\documentclass[%
aip,
amsmath,amssymb,
reprint,%
]{revtex4-1}

\usepackage{graphicx}
\usepackage{dcolumn}
\usepackage{bm}

\usepackage[utf8]{inputenc}
\usepackage[T1]{fontenc}
\usepackage{mathptmx}
\usepackage{color}
\begin{document}
		
\title{Magnetization reversal of dipolar coupled nanomagnets studied by two-dimensional electron gas based micro-Hall magnetometry}

\author{N. Keswani}
\affiliation{Department of Physics, Indian Institute of Technology, Delhi, Hauz Khas, New Delhi 110016, India}

\author{Y. Nakajima}
\affiliation{Bio-Nano Electronics Research Centre, Toyo University, Kawagoe, Saitama 3508585, Japan}

\author{ N. Chauhan}
\affiliation{Bio-Nano Electronics Research Centre, Toyo University, Kawagoe, Saitama 3508585, Japan}

\author{H. Chakraborti}
\affiliation{Department of Physics, Indian Institute of Technology, Bombay, Mumbai 400076, India}

\author{K. Dasgupta}
\affiliation{Department of Physics, Indian Institute of Technology, Bombay, Mumbai 400076, India}

\author{S. Kumar}
\affiliation{Bio-Nano Electronics Research Centre, Toyo University, Kawagoe, Saitama 3508585, Japan}

\author{Y. Ohno$^!$}
\affiliation{Research Institute of Electrical Communication, Tohoku University, Sendai 980-8577, Japan}

\author{H. Ohno}
\affiliation{Research Institute of Electrical Communication, Tohoku University, Sendai 980-8577, Japan}

\author{P. Das}
\email{pintu@physics.iitd.ac.in}
\affiliation{Department of Physics, Indian Institute of Technology, Delhi, Hauz Khas, New Delhi 110016, India}

\date{\today}
	
	\begin{abstract}
		We report here the results of two-dimensional electron gas based micro-Hall magnetometry measurements and micromagnetic simulations of dipolar coupled nanomagnets of Ni$_{80}$Fe$_{20}$ arranged in a double ring-like geometry. We observe that although magnetic force microscopy images exhibit single domain like magnetic states for the nanostructures, their reversal processes may undergo complex behavior. The details of such reversal behavior is observed as specific features in micro-Hall magnetometry data which compares well with the micromagnetic simulation data.  
	\end{abstract}

	\maketitle
	
In the field of nanoscale magnetism, patterned magnetic nanostructures makes it possible to study the geometry dependent novel magnetic behavior that may be useful for modern spintronic devices\cite{hoffmann2015opportunities}. Therefore, controlled fabrication of nanostructures by lithography techniques resulted in a flurry of research activities in the last decade\cite{fernandez2017three}. Recently, there has been a strong focus on dipolar interaction mediated magnetic behavior of nanostructures from both fundamental as well as potential application point of views\cite{wang2006artificial,zhang2014innovative,perrin2016extensive,loehr2016defect,keller2018direct,ostman2018interaction,farhan2019emergent,sivasubramani2019dipole}. Nanomagnets of strong shape anisotropy can mimic single macrospin which can act as binary switches due to their two stable states\cite{cowburn2002probing,rahm2005programmable,imre2006majority}. Although such nanomagnets contain large no. of spins interacting via strong exchange interaction, their net macrospin like behavior can be used as a building block of potential single spin-like logic circuitry operating at room temperature. Energetically, using such nanomagnets interacting predominantly via dipolar interactions is advantageous compared to solid state based logic circuits\cite{salahuddin2007interacting}. While several developments towards designing such practical devices have been attempted where the role of individual nanomagnetic states are exploited, recent developments in creating more complex structures makes it interesting to utilize the collective behavior of the nanomagnets for such practical applications. Such coupled nanomagnetic system based computational logic devices have advantage of non volatility in addition to their low-power requirement and have attracted attention of researchers to design novel nanomagnetic structures for this purpose\cite{arava2018computational,luo2019chirally}. It is clear that in order to realize potential application of such structures, engineering and control of the magnetic states of such structures are essential. This requires an indepth understanding of the switching behavior of the nanomagnets in a dipolar coupled environment. Moreover, recent developments in the use of such shape anisotropic nanomagnets mimicking Ising spin-like behavior in understanding fundamentals of magnetic frustrations such as in artificial spin ice systems has opened new avenues to create arrays in different geometries and explore the underlying physics\cite{nisoli2013colloquium,nisoli2017deliberate,heyderman2013artificial,rougemaille2019cooperative}. While switching behavior of  nanomagnets of simple geometries has been adequately reported in literature, however, little is known about the exact switching behavior of such nanomagnets when placed in complex arrangements within dipolar coupled environment\cite{pohlit2016magnetic}. This is clearly needed in order to construct new geometries for fundamental studies\cite{nisoli2013colloquium,stamps2014artificial,gilbert2016emergent,nisoli2017deliberate} as well as applications~\cite{sivasubramani2019dipole,arava2019engineering}. Additionally, strong shape anisotropic nanomagnets, which are most often considered as exhibiting Ising spin-like behavior, may be effected by the unintentional defects introduced during fabrication. Thus, it is essential to investigate how such defects modify the local magnetic behavior which in turn may play a significant role in the overall switching behavior of the corresponding nanostructures. 
\begin{figure}
\includegraphics[width=1\linewidth]{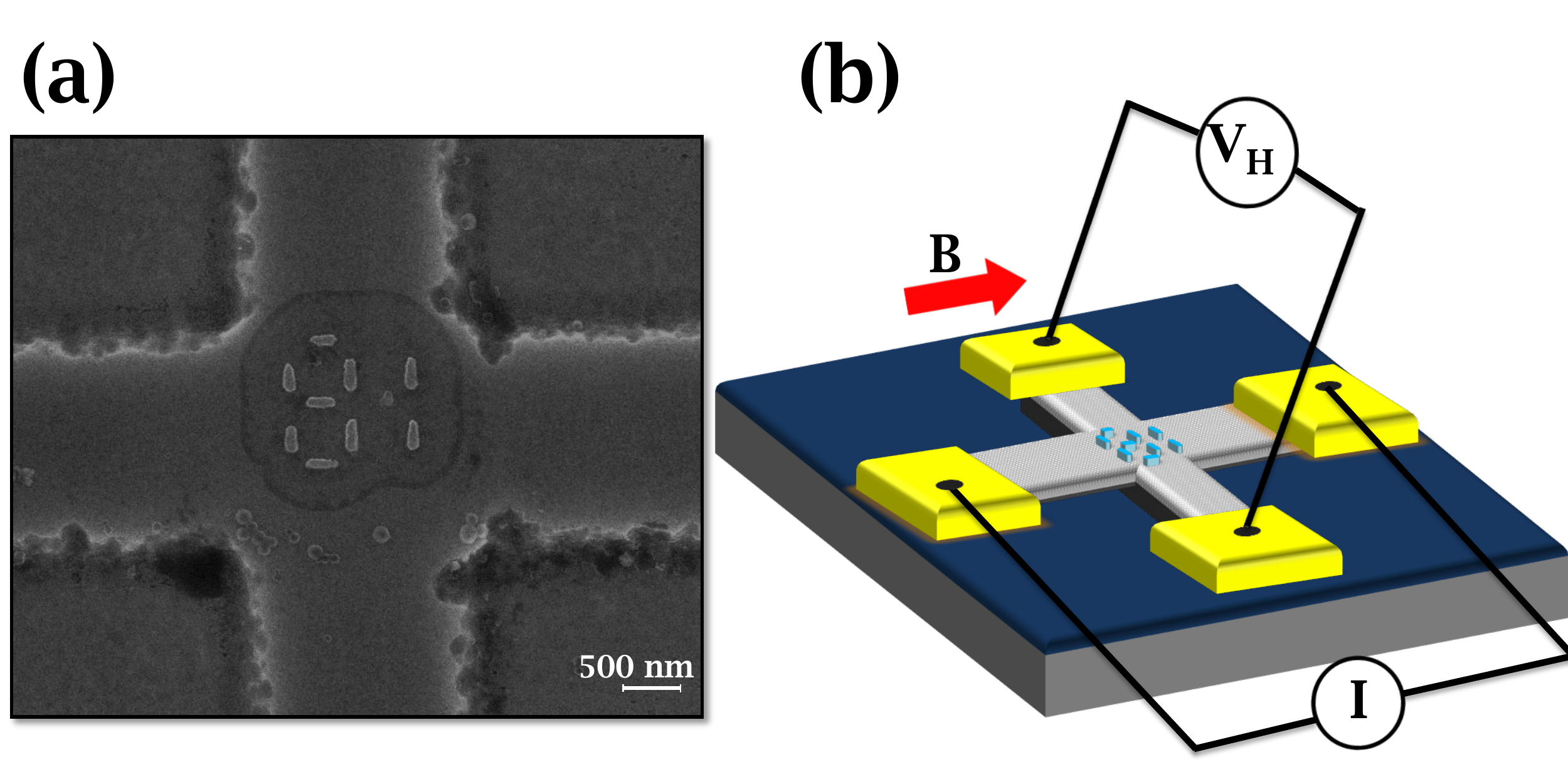}
\caption{\label{Fig1} (a) SEM image of nanomagnets grown on GaAs/AlGaAs Hall sensor. (b) Schematic diagram of the measurement set-up.}
\end{figure}
In this work, we investigated the collective switching behavior of nine dipolar coupled nanomagnets of strong shape anisotropy arranged in two ring-like geometry. Such rings are building blocks of different engineered systems and therefore, understanding their switching behavior in this geometry may be helpful to elucidate their behavior in more complex structures. In our earlier work of micromagnetic simulations, we observed simultaneous magnetization reversals of nanomagnets which are in the next nearest neighbor positions suggesting an indirect coupling of the nanomagnets in the given system~\cite{keswani2019micromagnetic}. 
In order to understand the role of the next nearest neighbor nanomagnetic elements on the ring structure, two nanomagnets of same dimensions as the others were patterned at next nearest neighbor positions without changing the symmetry as shown in Fig.\,\ref{Fig1}(a). Although our magnetic force microscopy (MFM) measurements show that all these nanomagnets are in single domain state, however, detailed magnetization reversals studied by employing highly sensitive two-dimensional electron gas (2-DEG) based micro-Hall magnetometry technique reveal several interesting features. Micromagnetic simulations show that these features are due to changes in the magnetic state of individual nanomagnets in the dipolar coupled environment. 

For our studies, an array of Hall devices was fabricated from a molecular beam epitaxy (MBE)-grown modulation-doped GaAs/AlGaAs heterostructure\cite{luo1988low}. The 2-DEG of the heterostructure lies approximately 230\,nm below the surface. Hall bars of 2$\times 2\,\mu$m$^2$  are patterned using electron-beam lithography (EBL) followed by wet-chemical etching. Ohmic contacts with the 2-DEG were ensured by following a metallization steps involving Ni, Au and Ge layers~\cite{goktacs2008alloyed}. The detailed fabrication steps are discussed in our earlier work\cite{keswani2018fabrication}. Shape anisotropic nanomagnets of Ni$_{80}$Fe$_{20}$ of dimensions 300$\times$100$\times$25\,nm$^3$ are defined on the active area of Hall bars using a second EBL step in combination with lift-off process (see Fig.\,\ref{Fig1}). 
The center-to-center distance between each nanoisland is 450\,nm. A Ti layer of thickness 5\,nm was used to increase the adhesion of permalloy on the GaAs surface. A capping layer of Al of thickness 5\,nm was deposited on Ni$_{80}$Fe$_{20}$ to prevent oxidation of the magnetic layer. Entire deposition was carried out using electron beam induced deposition without breaking the vacuum. For measurements, the Hall sensors were wire bonded to a leadless chip carrier. The measurements were carried out in an oxford instruments' Heliox cryostat.  
The sheet carrier density(\textit{n}) and mobility($\mu$) of the 2-DEG were determined to be 3.59$\times$10$^{11}$/cm$^{2}$ and 3.7$\times$ 10$^{5}$ cm$^2$/Vs, respectively at $T=300$\,mK. The magnetic measurements were carried out at $T=1.6$\,K with external magnetic field applied in plane as shown in Fig.\,\ref{Fig1}(b). In this measurements geometry, the measured Hall voltage ($V_{\rm{H}}$) is proportional to the average z-component of the magnetic stray field ($<B_{\rm{z}}>$) emanating from the magnetic nanostructures. This was confirmed from the Hall voltage measurements on an empty Hall cross (not shown). The measurements were carried out using the standard Lock-In technique.  
\begin{figure}
	\includegraphics[width=0.5\textwidth]{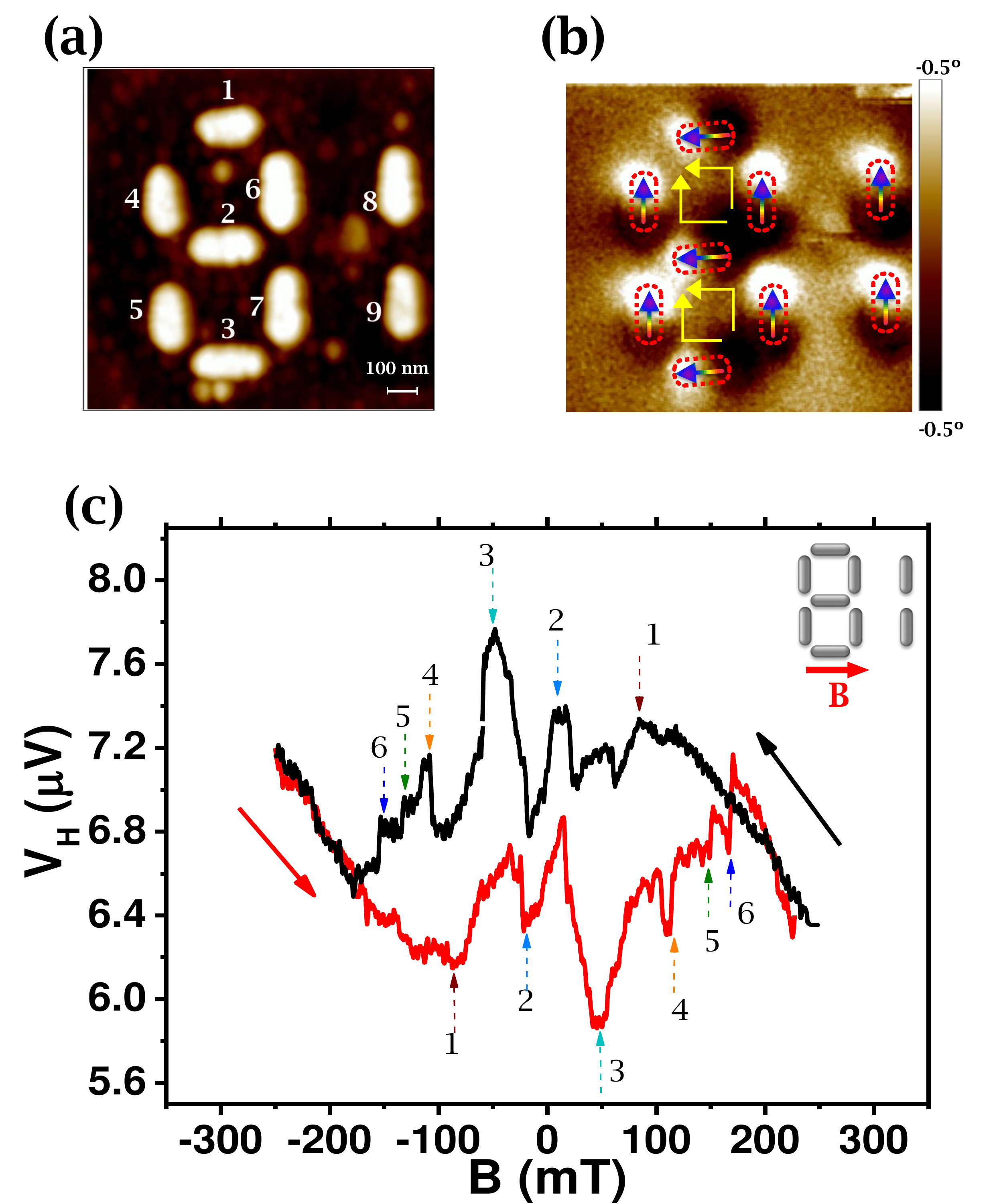}
	\caption{\label{Fig2}(a) AFM image of the magnetic nanoislands at room temperature. (b) Corresponding MFM image showing magnetic contrasts of single domain islands of Ni$_{80}$Fe$_{20}$. Dotted loops with arrows are guide to the eyes for the magnetization directions of the nanoislands. Magnetization of the rings forming onion type lops are shown by four other arrows.(c) Micro-Hall magnetometry data showing hysteresis in Hall voltage of 2-DEG versus external magnetic field. The features in the hystersis loops are identified by numbers and arrows. Upsweep and downsweep features at corresponding fields are identified by the same numbers. Inset shows the schematics of the nanostructure and applied field direction.}
\end{figure}

Fig.\,\ref{Fig2}(a) shows the topography of the nanostructures as obtained using atomic force microscope (AFM). Irregularities in the shapes of the nanoislands are observed which arises from the weak adhesion of Ni$_{80}$Fe$_{20}$ on GaAs surface as well as the lift-off process. However, the corresponding magnetic force microscopy (MFM) data at the remanent state exhibit clear bright and dark patches in each nanomagnet showing that all the nanomagnets are magnetically in single domain state (see Fig.\ref{Fig2}(b)). As mentioned above, such single domain state can be considered as classical analogue of Ising like macrospin and have been exploited in ASI or logic devices. Moreover, the data show that the three horizontal as well as six vertically placed nanomagnets are ferromagnetically aligned. Such ferromagnetically aligned nanomagnets in two different sublattices form onion type loops in the two ring-like arrangements which is evident from Fig.\ref{Fig2}(b) ~\cite{keswani2018magnetization}. The MFM data further indicate that the two nanomagnets at next nearest neighbor positions do not interact significantly with the elements involved in the ring structure. This is evidenced by the observation of ferromagnetic alignment of the corresponding vertical nanomagnets.


Next, in order to investigate the detailed switching behavior of the dipolarly coupled nanomagnets, measurements of <$B_z$> of the nanomagnetic system were carried out using high-sensitive micro-Hall magnetometry method. The measurements were carried out at $T=1.6$\,K where the electronic transport in the 2-DEG occurs in the ballistic regime. Due to the strong shape anisotropy ($K_s\sim 6.5\times 10^4$Jm$^{-3}$) of these nanomagnets, the nanomagnets are athermal. 
Saturation field ($B_{\rm{ext}}$) for these nanomagnets is $\sim200$\,mT. The measurements were carried out for in-plane external field of $\pm300$\,mT applied along [10] direction which is the easy axis for the horizontally placed nanomagnetic islands (hard axis for the vertical islands, see inset of Fig.\,\ref{Fig2}(c)). As the field is swept within $\pm$\,300\,mT, a hysteresis is observed in the Hall voltage of the 2-DEG with several reproducible features for both down and up sweep of the field, as shown in Fig.\,\ref{Fig2}(c). These changes in the Hall voltage reflects the changes in the magnetic state of the dipolarly coupled nanomagnets. The data exhibit distinct steps as well as broad peaks at specific field values while sweeping the field along both the directions. Specifically, we observe the major features as a sharp drop of signal at $\sim\pm82$\,mT(1), three peaks at $\sim\pm15$\,mT(2), $\sim\mp50$\,mT(3), and  $\sim\mp 110$\,mT(4), respectively and two sharp jumps at $\sim\mp 125$\,mT(5) and $\sim\mp 150$\,mT(6), respectively. Here, the first sign is for downsweep and the second for upsweep field, respectively. The numbers in brackets are to identify the features as indicated also by arrows in Fig.\,\ref{Fig2}(c). Interestingly, two features as described above are observed in the first quadrant of $V_H$-$B$ loop. 

\begin{figure}
 	\includegraphics[width=0.51\textwidth]{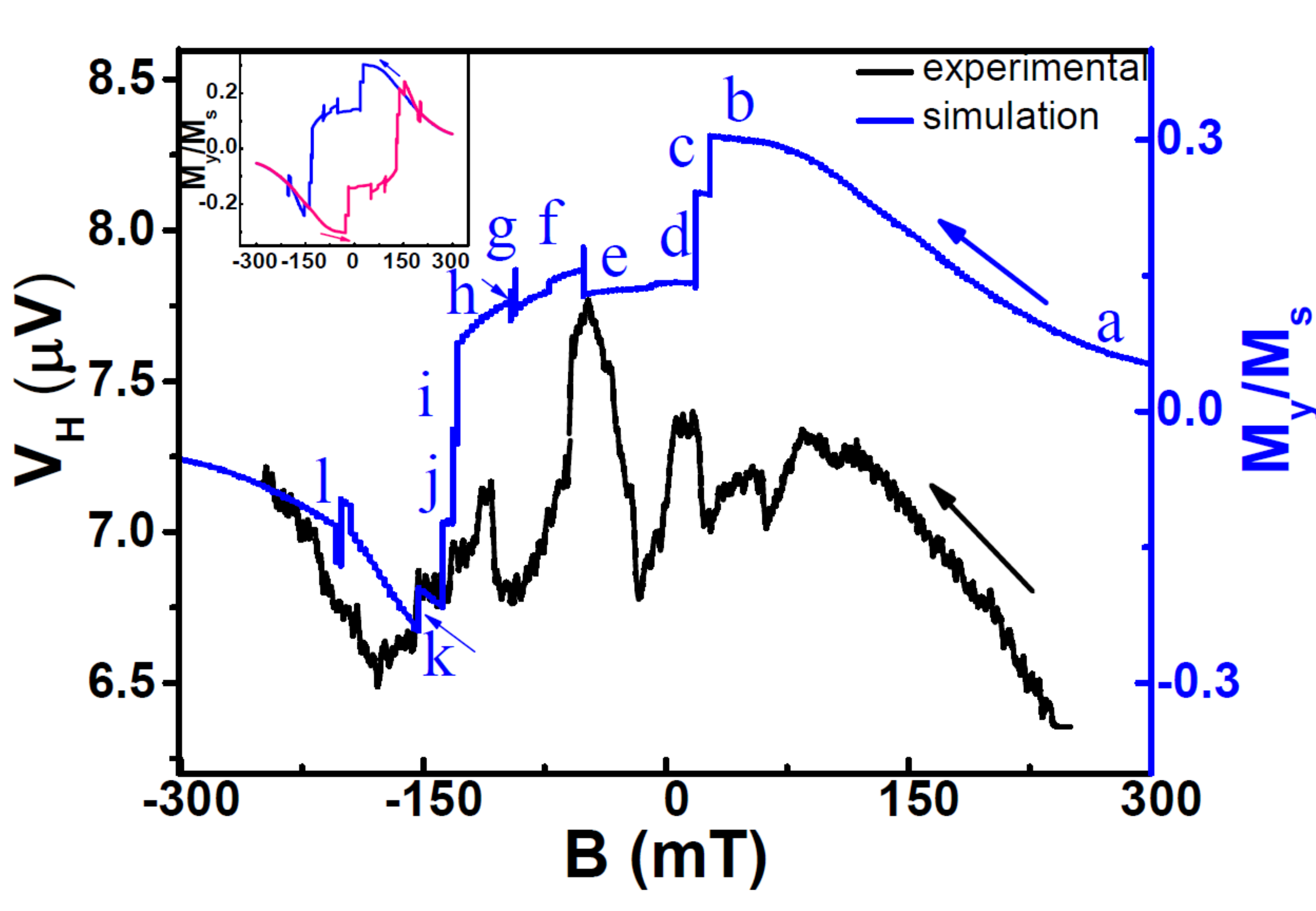}
\caption{\label{Fig3} Magnetization $M_y$ as obtained from simulation as well as Hall voltage $V_H$ due to average z-component of stray field <$B_z$> measured in the Hall voltage of 2-DEG. Inset shows the complete hysteresis loop obtained from simulation.}
 \end{figure}

While the sharp jumps may indicate magnetization switchings of individual nanoislands, the origin of the other reproducible features such as peaks appearing in both field sweep directions are not immediately clear\cite{pohlit2016magnetic}. These results may indicate specific changes occurring within the coupled nanostructures induced by external field which is not directly accessible by MFM measurements. We note here that the high sensitivity of the 2-DEG based Hall sensors have been used to detect nucleation and annihilation of magnetic vortices in individual nanostructures, interaction of domain walls with Peirls potential,etc.\cite{novoselov2003subatomic,lipert2010individual,das2010magnetization,matsunaga2013detection}. In order to understand the observed features which are likely to be the results of complex switching processes involving the multiple nanostructures, we carried out micromagnetic simulations of the entire system in presence of in-plane field applied along [10] direction. Our ground state simulations were performed using finite difference based Object Oriented Micro Magnetic Framework (OOMMF)~\cite{donahue1999national}. Typical experimentally reported values of saturation magnetization $M_s =$  8.6 $\times$10$^5$\,A/m, the exchange stiffness constant $A = $13\,pJ/m and damping constant of 0.5 for Ni$_{80}$Fe$_{20}$ are used for the calculations~\cite{principles}. The magnetocrystalline anisotropy is neglected in the computation. For the simulations, we used the exact experimental structure of the nanomagnets as obtained by AFM.

\begin{figure}

 	\includegraphics[width=0.92\linewidth]{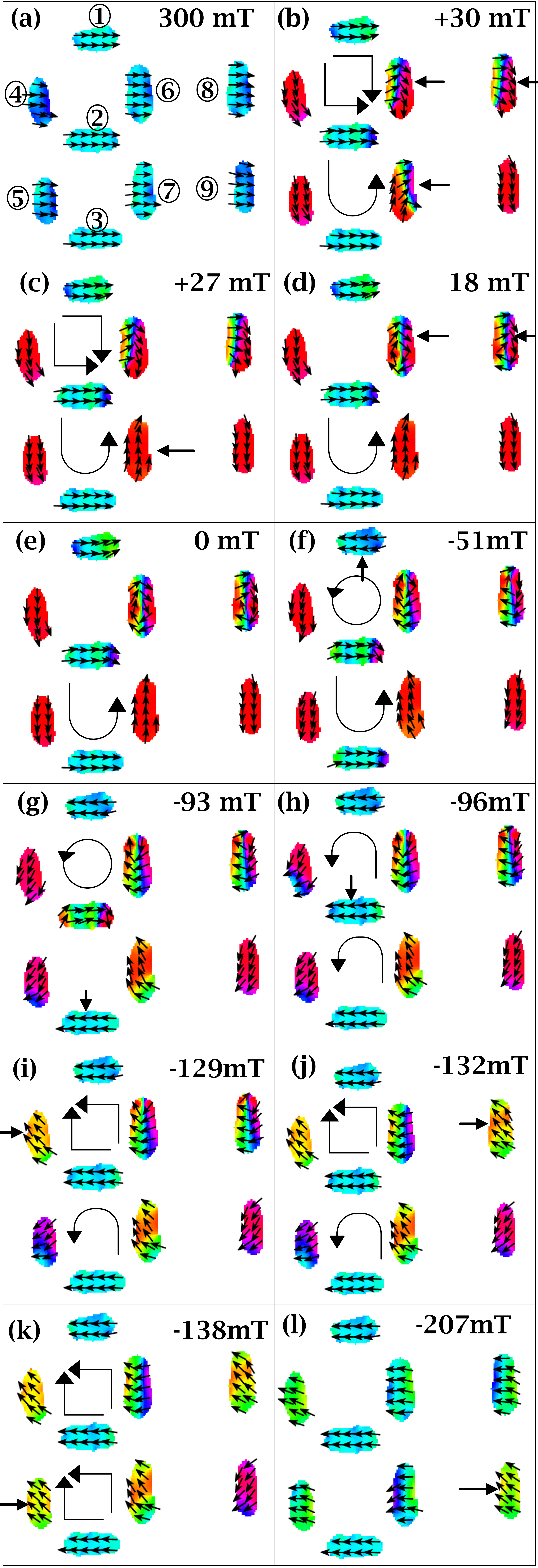}
\caption{\label{Fig4} Micromagnetic states of the interacting nanostructures at different magnetic field values corresponing to the features observed in the hysteresis loop shown in Fig.\ref{Fig3}. The nos. in (a) are to identify the nanostructures. The arrows refer to the magnetization switchings of the nanomagnets at the corresponding external fields.}

 \end{figure}

For the nanomagnetic system under consideration, any deviation from single domain Ising-like state is reflected in the magnetization $M_y$ along the [01] direction. 

The inset of Fig.\,\ref{Fig3} shows the simulated results of hysteresis in $M_y$ for the nanostructures. For comparison, simulation data as well as the experimental data for only downsweep field are plotted in the main figure. As can be seen from the plot, the simulation results captures several experimentally measured features remarkably well. Particularly, we observe clear features due to the magnetic activity before remanence, the peak at about $\mp 51$\,mT and several other sharp jumps in the field range where features in the experimental data are observed. 
The results allow us to investigate in details the exact micromagnetic state of individual nanomagnets and the changes of these states in the dipolarly coupled environment which is induced by the external field.

At the first quadrant of the $M_y$ vs $B$ loop (i.e., before remanence), simulation results exhibit features at 30\,mT, 27\,mT and 18\,mT, respectively. As the field is downsweeped from the saturation, we observe that the detailed micromagnetic behavior of some of the nanostructures changes differently which is most likely due to the irregularities in the structures and the different local dipolar fields which the nanostructures experience. These micromagnetic analysis are shown in Fig.\,\ref{Fig4}. For $B<B_{\rm{sat}}$, the magnetization of the nanomagnets 4, 5 and 9  (as identified in Fig.\,\ref{Fig2}) starts to rotate towards their easy axes whereas 6 and 8 start to form vortices (see the discussions below). The magnet 7 behaves differently which, at about 27\,mT, suddenly switches stabilizing a horse-shoe type loop at the lower ring whereas the upper ring shows an onion-type loop as shown in Fig.\,\ref{Fig4}(c). At about 18\,mT, both nanomagnets 6 and 8 forms vortices where features are observed both in $M_y$ as well as experimentally measured $V_H$ (see Fig.\,\ref{Fig4}(d)). Between $\sim 18$\,mT and -50\,mT, the vortex cores for magnets 6 and 8 appears to shift from lower end to the upper end. At about -51\,mT, the island 1 switches thus forming a microvortex state at the upper ring. Therefore, the upper ring has a microvortex and lower ring has a horse-shoe state as seen in Fig.\,\ref{Fig4}(f). At this field, a peak is observed in the experiments and a corresponding reverse step is observed for $M_y$. At -93\,mT, magnetization of the island 3 switches thus converting the lower ring to two head-to-head and two tail-to-tail type loop (see Fig.\,\ref{Fig4}(g)). At $\sim$-96\,mT, island 2 switches thereby forming a horse-shoe type state for both the rings as shown in Fig.\,\ref{Fig4}(h). These changes in $M$ appear to result in to a peak in the average <$B_z$> which is observed between -97\,mT and -110\,mT in the measured Hall voltage. With the switching of nanomagnet 2, the magnetizations of all the horizontal islands align along the applied field direction. As the field reaches $\sim$-129\,mT, the island 4 undergoes a sudden change in magnetization orientation showing a switching like behavior, thus horse-shoe state in upper ring changes to onion state (see Fig.\,\ref{Fig4}(i)). In general, such sudden change in magnetization for this nanomagnet is unexpected as the applied field direction is along the hard axis. A corresponding sharp step in $V_H$ vs $B$ data is observed at this field. At $\sim$ -132\,mT, the magnetization of island 8 suddenly changes, exhibiting a switching-like behavior. Next, the island 5 switches at -138\,mT. This reversal converts horse-shoe state to onion state in the lower ring, which forms onion states in both the rings. Finally, at -207\,mT, the magnetization of the vertical nanoisland 9 switches and thereby saturating the magnetization. It is interesting to note that with the applied external field direction in this case, which is along the hard direction for the vertically placed nanomagnets, gradual rotation of magnetization and not sudden switching in individual nanomagnets are expected for magnets 4,5 and 8 and 9. However, clearly, these nanomagnets show sudden switching-like behavior demonstrating that the dipolar interaction may lead to nontrivial micromagnetic states in nanomagnetic systems which are otherwise in a single domain state as also observed from MFM data as shown in Fig.\ref{Fig2}(b).
To determine if there is any role of dipolar interaction in stabilizing magnetic votex states  in the nanomagnets 6 and 8, micromagnetic states of these nanostructures in isolated state are investigated. It is observed that for applied field along the hard direction, the nanomagnets indeed exhibit magnetic vortex states without any dipolar interaction (not shown). However, the nanomagnet 4 shows single domain state at remanence.  This is observed in the dipolar coupled environment as well. This shows that the vortices as observed in the two nanomagnets may be due to the specific geometry of the nanomagnets. The experimental and simulation results clearly show that although the nanomagnets show single domain like behavior, the real structures may undergo complex switching processes which do not confirm to Stoner- Wohlfahrth like behavior as is known for ideal single domain magnetic states. 

In conclusion, we have investigated the magnetization reversal behavior in dipolarly coupled nanomagnets forming two coupled ring-like structures. The remanent state as observed using MFM show that all the nanomagnets are in single domain state. However, detailed (continuous) field dependent high-sensitive measurements of average stray fields emanating from the nanomagnets show features which can be identified as due to specific micromagnetic states in the nanomagnets. This is explained with the micromagnetic simulation results which match reasonably well with the experimental data. The results suggest that local irregularities affect the exact micromagnetic state thereby converting two islands of single domain dimensions to a magnetically vortex state. Our  results also demonstrate the remarkable ability of 2-DEG based micro-Hall magnetometry method used in ballistic transport regime in detecting changes in stray fields due to local micromagnetic changes. 

\begin{acknowledgments}
We acknowledge Tomofumi Ukai for valuable discussions. We are thankful to Yasuhiko Fujii and Masahide Tokuda for help in fabrication of Hall devices. N.K. is also thankful to University Grant Commission, (UGC) Govt. of India, for providing research fellowship. 
\end{acknowledgments}
! Present address: Faculty of Pure and Applied Physics, University of Tsukuba, Japan.

\bibliography{ref}

\begin{thebibliography}{35}%
\makeatletter
\providecommand \@ifxundefined [1]{%
 \@ifx{#1\undefined}
}%
\providecommand \@ifnum [1]{%
 \ifnum #1\expandafter \@firstoftwo
 \else \expandafter \@secondoftwo
 \fi
}%
\providecommand \@ifx [1]{%
 \ifx #1\expandafter \@firstoftwo
 \else \expandafter \@secondoftwo
 \fi
}%
\providecommand \natexlab [1]{#1}%
\providecommand \enquote  [1]{``#1''}%
\providecommand \bibnamefont  [1]{#1}%
\providecommand \bibfnamefont [1]{#1}%
\providecommand \citenamefont [1]{#1}%
\providecommand \href@noop [0]{\@secondoftwo}%
\providecommand \href [0]{\begingroup \@sanitize@url \@href}%
\providecommand \@href[1]{\@@startlink{#1}\@@href}%
\providecommand \@@href[1]{\endgroup#1\@@endlink}%
\providecommand \@sanitize@url [0]{\catcode `\\12\catcode `\$12\catcode
  `\&12\catcode `\#12\catcode `\^12\catcode `\_12\catcode `\%12\relax}%
\providecommand \@@startlink[1]{}%
\providecommand \@@endlink[0]{}%
\providecommand \url  [0]{\begingroup\@sanitize@url \@url }%
\providecommand \@url [1]{\endgroup\@href {#1}{\urlprefix }}%
\providecommand \urlprefix  [0]{URL }%
\providecommand \Eprint [0]{\href }%
\providecommand \doibase [0]{http://dx.doi.org/}%
\providecommand \selectlanguage [0]{\@gobble}%
\providecommand \bibinfo  [0]{\@secondoftwo}%
\providecommand \bibfield  [0]{\@secondoftwo}%
\providecommand \translation [1]{[#1]}%
\providecommand \BibitemOpen [0]{}%
\providecommand \bibitemStop [0]{}%
\providecommand \bibitemNoStop [0]{.\EOS\space}%
\providecommand \EOS [0]{\spacefactor3000\relax}%
\providecommand \BibitemShut  [1]{\csname bibitem#1\endcsname}%
\let\auto@bib@innerbib\@empty
\bibitem [{\citenamefont {Hoffmann}\ and\ \citenamefont
  {Bader}(2015)}]{hoffmann2015opportunities}%
  \BibitemOpen
  \bibfield  {author} {\bibinfo {author} {\bibfnamefont {A.}~\bibnamefont
  {Hoffmann}}\ and\ \bibinfo {author} {\bibfnamefont {S.~D.}\ \bibnamefont
  {Bader}},\ }\href@noop {} {\bibfield  {journal} {\bibinfo  {journal} {Phys.
  Rev. Appl.}\ }\textbf {\bibinfo {volume} {4}},\ \bibinfo {pages} {047001}
  (\bibinfo {year} {2015})}\BibitemShut {NoStop}%
\bibitem [{\citenamefont {Fern{\'a}ndez-Pacheco}\ \emph
  {et~al.}(2017)\citenamefont {Fern{\'a}ndez-Pacheco}, \citenamefont
  {Streubel}, \citenamefont {Fruchart}, \citenamefont {Hertel}, \citenamefont
  {Fischer},\ and\ \citenamefont {Cowburn}}]{fernandez2017three}%
  \BibitemOpen
  \bibfield  {author} {\bibinfo {author} {\bibfnamefont {A.}~\bibnamefont
  {Fern{\'a}ndez-Pacheco}}, \bibinfo {author} {\bibfnamefont {R.}~\bibnamefont
  {Streubel}}, \bibinfo {author} {\bibfnamefont {O.}~\bibnamefont {Fruchart}},
  \bibinfo {author} {\bibfnamefont {R.}~\bibnamefont {Hertel}}, \bibinfo
  {author} {\bibfnamefont {P.}~\bibnamefont {Fischer}}, \ and\ \bibinfo
  {author} {\bibfnamefont {R.~P.}\ \bibnamefont {Cowburn}},\ }\href@noop {}
  {\bibfield  {journal} {\bibinfo  {journal} {Nature Comm.}\ }\textbf {\bibinfo
  {volume} {8}},\ \bibinfo {pages} {15756} (\bibinfo {year}
  {2017})}\BibitemShut {NoStop}%
\bibitem [{\citenamefont {Wang}\ \emph {et~al.}(2006)\citenamefont {Wang},
  \citenamefont {Nisoli}, \citenamefont {Freitas}, \citenamefont {Li},
  \citenamefont {McConville}, \citenamefont {Cooley}, \citenamefont {Lund},
  \citenamefont {Samarth}, \citenamefont {Leighton}, \citenamefont {Crespi}
  \emph {et~al.}}]{wang2006artificial}%
  \BibitemOpen
  \bibfield  {author} {\bibinfo {author} {\bibfnamefont {R.}~\bibnamefont
  {Wang}}, \bibinfo {author} {\bibfnamefont {C.}~\bibnamefont {Nisoli}},
  \bibinfo {author} {\bibfnamefont {R.~S.~d.}\ \bibnamefont {Freitas}},
  \bibinfo {author} {\bibfnamefont {J.}~\bibnamefont {Li}}, \bibinfo {author}
  {\bibfnamefont {W.}~\bibnamefont {McConville}}, \bibinfo {author}
  {\bibfnamefont {B.}~\bibnamefont {Cooley}}, \bibinfo {author} {\bibfnamefont
  {M.}~\bibnamefont {Lund}}, \bibinfo {author} {\bibfnamefont {N.}~\bibnamefont
  {Samarth}}, \bibinfo {author} {\bibfnamefont {C.}~\bibnamefont {Leighton}},
  \bibinfo {author} {\bibfnamefont {V.}~\bibnamefont {Crespi}},  \emph
  {et~al.},\ }\href@noop {} {\bibfield  {journal} {\bibinfo  {journal}
  {Nature}\ }\textbf {\bibinfo {volume} {439}},\ \bibinfo {pages} {303}
  (\bibinfo {year} {2006})}\BibitemShut {NoStop}%
\bibitem [{\citenamefont {Zhang}\ \emph {et~al.}(2014)\citenamefont {Zhang},
  \citenamefont {Yang}, \citenamefont {Wang},\ and\ \citenamefont
  {Zhang}}]{zhang2014innovative}%
  \BibitemOpen
  \bibfield  {author} {\bibinfo {author} {\bibfnamefont {B.}~\bibnamefont
  {Zhang}}, \bibinfo {author} {\bibfnamefont {X.}~\bibnamefont {Yang}},
  \bibinfo {author} {\bibfnamefont {Z.}~\bibnamefont {Wang}}, \ and\ \bibinfo
  {author} {\bibfnamefont {M.}~\bibnamefont {Zhang}},\ }\href@noop {}
  {\bibfield  {journal} {\bibinfo  {journal} {Micro \& Nano Lett.}\ }\textbf
  {\bibinfo {volume} {9}},\ \bibinfo {pages} {359--362} (\bibinfo {year}
  {2014})}\BibitemShut {NoStop}%
\bibitem [{\citenamefont {Perrin}, \citenamefont {Canals},\ and\ \citenamefont
  {Rougemaille}(2016)}]{perrin2016extensive}%
  \BibitemOpen
  \bibfield  {author} {\bibinfo {author} {\bibfnamefont {Y.}~\bibnamefont
  {Perrin}}, \bibinfo {author} {\bibfnamefont {B.}~\bibnamefont {Canals}}, \
  and\ \bibinfo {author} {\bibfnamefont {N.}~\bibnamefont {Rougemaille}},\
  }\href@noop {} {\bibfield  {journal} {\bibinfo  {journal} {Nature}\ }\textbf
  {\bibinfo {volume} {540}},\ \bibinfo {pages} {410} (\bibinfo {year}
  {2016})}\BibitemShut {NoStop}%
\bibitem [{\citenamefont {Loehr}, \citenamefont {Ortiz-Ambriz},\ and\
  \citenamefont {Tierno}(2016)}]{loehr2016defect}%
  \BibitemOpen
  \bibfield  {author} {\bibinfo {author} {\bibfnamefont {J.}~\bibnamefont
  {Loehr}}, \bibinfo {author} {\bibfnamefont {A.}~\bibnamefont {Ortiz-Ambriz}},
  \ and\ \bibinfo {author} {\bibfnamefont {P.}~\bibnamefont {Tierno}},\
  }\href@noop {} {\bibfield  {journal} {\bibinfo  {journal} {Phys. Rev. Lett.}\
  }\textbf {\bibinfo {volume} {117}},\ \bibinfo {pages} {168001} (\bibinfo
  {year} {2016})}\BibitemShut {NoStop}%
\bibitem [{\citenamefont {Keller}\ \emph {et~al.}(2018)\citenamefont {Keller},
  \citenamefont {Al~Mamoori}, \citenamefont {Pieper}, \citenamefont {Gspan},
  \citenamefont {Stockem}, \citenamefont {Schr{\"o}der}, \citenamefont {Barth},
  \citenamefont {Winkler}, \citenamefont {Plank}, \citenamefont {Pohlit} \emph
  {et~al.}}]{keller2018direct}%
  \BibitemOpen
  \bibfield  {author} {\bibinfo {author} {\bibfnamefont {L.}~\bibnamefont
  {Keller}}, \bibinfo {author} {\bibfnamefont {M.~K.}\ \bibnamefont
  {Al~Mamoori}}, \bibinfo {author} {\bibfnamefont {J.}~\bibnamefont {Pieper}},
  \bibinfo {author} {\bibfnamefont {C.}~\bibnamefont {Gspan}}, \bibinfo
  {author} {\bibfnamefont {I.}~\bibnamefont {Stockem}}, \bibinfo {author}
  {\bibfnamefont {C.}~\bibnamefont {Schr{\"o}der}}, \bibinfo {author}
  {\bibfnamefont {S.}~\bibnamefont {Barth}}, \bibinfo {author} {\bibfnamefont
  {R.}~\bibnamefont {Winkler}}, \bibinfo {author} {\bibfnamefont
  {H.}~\bibnamefont {Plank}}, \bibinfo {author} {\bibfnamefont
  {M.}~\bibnamefont {Pohlit}},  \emph {et~al.},\ }\href@noop {} {\bibfield
  {journal} {\bibinfo  {journal} {Sci. Rep.}\ }\textbf {\bibinfo {volume}
  {8}},\ \bibinfo {pages} {6160} (\bibinfo {year} {2018})}\BibitemShut
  {NoStop}%
\bibitem [{\citenamefont {{\"O}stman}\ \emph {et~al.}(2018)\citenamefont
  {{\"O}stman}, \citenamefont {Stopfel}, \citenamefont {Chioar}, \citenamefont
  {Arnalds}, \citenamefont {Stein}, \citenamefont {Kapaklis},\ and\
  \citenamefont {Hj{\"o}rvarsson}}]{ostman2018interaction}%
  \BibitemOpen
  \bibfield  {author} {\bibinfo {author} {\bibfnamefont {E.}~\bibnamefont
  {{\"O}stman}}, \bibinfo {author} {\bibfnamefont {H.}~\bibnamefont {Stopfel}},
  \bibinfo {author} {\bibfnamefont {I.-A.}\ \bibnamefont {Chioar}}, \bibinfo
  {author} {\bibfnamefont {U.~B.}\ \bibnamefont {Arnalds}}, \bibinfo {author}
  {\bibfnamefont {A.}~\bibnamefont {Stein}}, \bibinfo {author} {\bibfnamefont
  {V.}~\bibnamefont {Kapaklis}}, \ and\ \bibinfo {author} {\bibfnamefont
  {B.}~\bibnamefont {Hj{\"o}rvarsson}},\ }\href@noop {} {\bibfield  {journal}
  {\bibinfo  {journal} {Nature Phys.}\ }\textbf {\bibinfo {volume} {14}},\
  \bibinfo {pages} {375} (\bibinfo {year} {2018})}\BibitemShut {NoStop}%
\bibitem [{\citenamefont {Farhan}\ \emph {et~al.}(2019)\citenamefont {Farhan},
  \citenamefont {Saccone}, \citenamefont {Petersen}, \citenamefont {Dhuey},
  \citenamefont {Chopdekar}, \citenamefont {Huang}, \citenamefont {Kent},
  \citenamefont {Chen}, \citenamefont {Alava}, \citenamefont {Lippert} \emph
  {et~al.}}]{farhan2019emergent}%
  \BibitemOpen
  \bibfield  {author} {\bibinfo {author} {\bibfnamefont {A.}~\bibnamefont
  {Farhan}}, \bibinfo {author} {\bibfnamefont {M.}~\bibnamefont {Saccone}},
  \bibinfo {author} {\bibfnamefont {C.~F.}\ \bibnamefont {Petersen}}, \bibinfo
  {author} {\bibfnamefont {S.}~\bibnamefont {Dhuey}}, \bibinfo {author}
  {\bibfnamefont {R.~V.}\ \bibnamefont {Chopdekar}}, \bibinfo {author}
  {\bibfnamefont {Y.-L.}\ \bibnamefont {Huang}}, \bibinfo {author}
  {\bibfnamefont {N.}~\bibnamefont {Kent}}, \bibinfo {author} {\bibfnamefont
  {Z.}~\bibnamefont {Chen}}, \bibinfo {author} {\bibfnamefont {M.~J.}\
  \bibnamefont {Alava}}, \bibinfo {author} {\bibfnamefont {T.}~\bibnamefont
  {Lippert}},  \emph {et~al.},\ }\href@noop {} {\bibfield  {journal} {\bibinfo
  {journal} {Sci. Adv.}\ }\textbf {\bibinfo {volume} {5}},\ \bibinfo {pages}
  {eaav6380} (\bibinfo {year} {2019})}\BibitemShut {NoStop}%
\bibitem [{\citenamefont {Sivasubramani}\ \emph {et~al.}(2019)\citenamefont
  {Sivasubramani}, \citenamefont {Mattela}, \citenamefont {Pal},\ and\
  \citenamefont {Acharyya}}]{sivasubramani2019dipole}%
  \BibitemOpen
  \bibfield  {author} {\bibinfo {author} {\bibfnamefont {S.}~\bibnamefont
  {Sivasubramani}}, \bibinfo {author} {\bibfnamefont {V.}~\bibnamefont
  {Mattela}}, \bibinfo {author} {\bibfnamefont {C.}~\bibnamefont {Pal}}, \ and\
  \bibinfo {author} {\bibfnamefont {A.}~\bibnamefont {Acharyya}},\ }\href@noop
  {} {\bibfield  {journal} {\bibinfo  {journal} {Nanotechnology}\ }\textbf
  {\bibinfo {volume} {31}},\ \bibinfo {pages} {025202} (\bibinfo {year}
  {2019})}\BibitemShut {NoStop}%
\bibitem [{\citenamefont {Cowburn}(2002)}]{cowburn2002probing}%
  \BibitemOpen
  \bibfield  {author} {\bibinfo {author} {\bibfnamefont {R.}~\bibnamefont
  {Cowburn}},\ }\href@noop {} {\bibfield  {journal} {\bibinfo  {journal} {Phys.
  Rev. B}\ }\textbf {\bibinfo {volume} {65}},\ \bibinfo {pages} {092409}
  (\bibinfo {year} {2002})}\BibitemShut {NoStop}%
\bibitem [{\citenamefont {Rahm}, \citenamefont {Stahl},\ and\ \citenamefont
  {Weiss}(2005)}]{rahm2005programmable}%
  \BibitemOpen
  \bibfield  {author} {\bibinfo {author} {\bibfnamefont {M.}~\bibnamefont
  {Rahm}}, \bibinfo {author} {\bibfnamefont {J.}~\bibnamefont {Stahl}}, \ and\
  \bibinfo {author} {\bibfnamefont {D.}~\bibnamefont {Weiss}},\ }\href@noop {}
  {\bibfield  {journal} {\bibinfo  {journal} {Appl. Phys. Lett.}\ }\textbf
  {\bibinfo {volume} {87}},\ \bibinfo {pages} {182107} (\bibinfo {year}
  {2005})}\BibitemShut {NoStop}%
\bibitem [{\citenamefont {Imre}\ \emph {et~al.}(2006)\citenamefont {Imre},
  \citenamefont {Csaba}, \citenamefont {Ji}, \citenamefont {Orlov},
  \citenamefont {Bernstein},\ and\ \citenamefont {Porod}}]{imre2006majority}%
  \BibitemOpen
  \bibfield  {author} {\bibinfo {author} {\bibfnamefont {A.}~\bibnamefont
  {Imre}}, \bibinfo {author} {\bibfnamefont {G.}~\bibnamefont {Csaba}},
  \bibinfo {author} {\bibfnamefont {L.}~\bibnamefont {Ji}}, \bibinfo {author}
  {\bibfnamefont {A.}~\bibnamefont {Orlov}}, \bibinfo {author} {\bibfnamefont
  {G.}~\bibnamefont {Bernstein}}, \ and\ \bibinfo {author} {\bibfnamefont
  {W.}~\bibnamefont {Porod}},\ }\href@noop {} {\bibfield  {journal} {\bibinfo
  {journal} {Science}\ }\textbf {\bibinfo {volume} {311}},\ \bibinfo {pages}
  {205--208} (\bibinfo {year} {2006})}\BibitemShut {NoStop}%
\bibitem [{\citenamefont {Salahuddin}\ and\ \citenamefont
  {Datta}(2007)}]{salahuddin2007interacting}%
  \BibitemOpen
  \bibfield  {author} {\bibinfo {author} {\bibfnamefont {S.}~\bibnamefont
  {Salahuddin}}\ and\ \bibinfo {author} {\bibfnamefont {S.}~\bibnamefont
  {Datta}},\ }\href@noop {} {\bibfield  {journal} {\bibinfo  {journal} {Appl.
  Phys. Lett.}\ }\textbf {\bibinfo {volume} {90}},\ \bibinfo {pages} {093503}
  (\bibinfo {year} {2007})}\BibitemShut {NoStop}%
\bibitem [{\citenamefont {Arava}\ \emph {et~al.}(2018)\citenamefont {Arava},
  \citenamefont {Derlet}, \citenamefont {Vijayakumar}, \citenamefont {Cui},
  \citenamefont {Bingham}, \citenamefont {Kleibert},\ and\ \citenamefont
  {Heyderman}}]{arava2018computational}%
  \BibitemOpen
  \bibfield  {author} {\bibinfo {author} {\bibfnamefont {H.}~\bibnamefont
  {Arava}}, \bibinfo {author} {\bibfnamefont {P.~M.}\ \bibnamefont {Derlet}},
  \bibinfo {author} {\bibfnamefont {J.}~\bibnamefont {Vijayakumar}}, \bibinfo
  {author} {\bibfnamefont {J.}~\bibnamefont {Cui}}, \bibinfo {author}
  {\bibfnamefont {N.~S.}\ \bibnamefont {Bingham}}, \bibinfo {author}
  {\bibfnamefont {A.}~\bibnamefont {Kleibert}}, \ and\ \bibinfo {author}
  {\bibfnamefont {L.~J.}\ \bibnamefont {Heyderman}},\ }\href@noop {} {\bibfield
   {journal} {\bibinfo  {journal} {Nanotechnology}\ }\textbf {\bibinfo {volume}
  {29}},\ \bibinfo {pages} {265205} (\bibinfo {year} {2018})}\BibitemShut
  {NoStop}%
\bibitem [{\citenamefont {Luo}\ \emph {et~al.}(2019)\citenamefont {Luo},
  \citenamefont {Dao}, \citenamefont {Hrabec}, \citenamefont {Vijayakumar},
  \citenamefont {Kleibert}, \citenamefont {Baumgartner}, \citenamefont {Kirk},
  \citenamefont {Cui}, \citenamefont {Savchenko}, \citenamefont {Krishnaswamy}
  \emph {et~al.}}]{luo2019chirally}%
  \BibitemOpen
  \bibfield  {author} {\bibinfo {author} {\bibfnamefont {Z.}~\bibnamefont
  {Luo}}, \bibinfo {author} {\bibfnamefont {T.~P.}\ \bibnamefont {Dao}},
  \bibinfo {author} {\bibfnamefont {A.}~\bibnamefont {Hrabec}}, \bibinfo
  {author} {\bibfnamefont {J.}~\bibnamefont {Vijayakumar}}, \bibinfo {author}
  {\bibfnamefont {A.}~\bibnamefont {Kleibert}}, \bibinfo {author}
  {\bibfnamefont {M.}~\bibnamefont {Baumgartner}}, \bibinfo {author}
  {\bibfnamefont {E.}~\bibnamefont {Kirk}}, \bibinfo {author} {\bibfnamefont
  {J.}~\bibnamefont {Cui}}, \bibinfo {author} {\bibfnamefont {T.}~\bibnamefont
  {Savchenko}}, \bibinfo {author} {\bibfnamefont {G.}~\bibnamefont
  {Krishnaswamy}},  \emph {et~al.},\ }\href@noop {} {\bibfield  {journal}
  {\bibinfo  {journal} {Science}\ }\textbf {\bibinfo {volume} {363}},\ \bibinfo
  {pages} {1435--1439} (\bibinfo {year} {2019})}\BibitemShut {NoStop}%
\bibitem [{\citenamefont {Nisoli}, \citenamefont {Moessner},\ and\
  \citenamefont {Schiffer}(2013)}]{nisoli2013colloquium}%
  \BibitemOpen
  \bibfield  {author} {\bibinfo {author} {\bibfnamefont {C.}~\bibnamefont
  {Nisoli}}, \bibinfo {author} {\bibfnamefont {R.}~\bibnamefont {Moessner}}, \
  and\ \bibinfo {author} {\bibfnamefont {P.}~\bibnamefont {Schiffer}},\
  }\href@noop {} {\bibfield  {journal} {\bibinfo  {journal} {Rev. Mod. Phys.}\
  }\textbf {\bibinfo {volume} {85}},\ \bibinfo {pages} {1473} (\bibinfo {year}
  {2013})}\BibitemShut {NoStop}%
\bibitem [{\citenamefont {Nisoli}, \citenamefont {Kapaklis},\ and\
  \citenamefont {Schiffer}(2017)}]{nisoli2017deliberate}%
  \BibitemOpen
  \bibfield  {author} {\bibinfo {author} {\bibfnamefont {C.}~\bibnamefont
  {Nisoli}}, \bibinfo {author} {\bibfnamefont {V.}~\bibnamefont {Kapaklis}}, \
  and\ \bibinfo {author} {\bibfnamefont {P.}~\bibnamefont {Schiffer}},\
  }\href@noop {} {\bibfield  {journal} {\bibinfo  {journal} {Nature Phys.}\
  }\textbf {\bibinfo {volume} {13}},\ \bibinfo {pages} {200} (\bibinfo {year}
  {2017})}\BibitemShut {NoStop}%
\bibitem [{\citenamefont {Heyderman}\ and\ \citenamefont
  {Stamps}(2013)}]{heyderman2013artificial}%
  \BibitemOpen
  \bibfield  {author} {\bibinfo {author} {\bibfnamefont {L.~J.}\ \bibnamefont
  {Heyderman}}\ and\ \bibinfo {author} {\bibfnamefont {R.~L.}\ \bibnamefont
  {Stamps}},\ }\href@noop {} {\bibfield  {journal} {\bibinfo  {journal} {J.
  Phys.: Condens. Matter}\ }\textbf {\bibinfo {volume} {25}},\ \bibinfo {pages}
  {363201} (\bibinfo {year} {2013})}\BibitemShut {NoStop}%
\bibitem [{\citenamefont {Rougemaille}\ and\ \citenamefont
  {Canals}(2019)}]{rougemaille2019cooperative}%
  \BibitemOpen
  \bibfield  {author} {\bibinfo {author} {\bibfnamefont {N.}~\bibnamefont
  {Rougemaille}}\ and\ \bibinfo {author} {\bibfnamefont {B.}~\bibnamefont
  {Canals}},\ }\href@noop {} {\bibfield  {journal} {\bibinfo  {journal} {Eur.
  Phys. J. B}\ }\textbf {\bibinfo {volume} {92}},\ \bibinfo {pages} {62}
  (\bibinfo {year} {2019})}\BibitemShut {NoStop}%
\bibitem [{\citenamefont {Pohlit}\ \emph {et~al.}(2016)\citenamefont {Pohlit},
  \citenamefont {Porrati}, \citenamefont {Huth}, \citenamefont {Ohno},
  \citenamefont {Ohno},\ and\ \citenamefont {M{\"u}ller}}]{pohlit2016magnetic}%
  \BibitemOpen
  \bibfield  {author} {\bibinfo {author} {\bibfnamefont {M.}~\bibnamefont
  {Pohlit}}, \bibinfo {author} {\bibfnamefont {F.}~\bibnamefont {Porrati}},
  \bibinfo {author} {\bibfnamefont {M.}~\bibnamefont {Huth}}, \bibinfo {author}
  {\bibfnamefont {Y.}~\bibnamefont {Ohno}}, \bibinfo {author} {\bibfnamefont
  {H.}~\bibnamefont {Ohno}}, \ and\ \bibinfo {author} {\bibfnamefont
  {J.}~\bibnamefont {M{\"u}ller}},\ }\href@noop {} {\bibfield  {journal}
  {\bibinfo  {journal} {J. Magn. Magn. Mater}\ }\textbf {\bibinfo {volume}
  {400}},\ \bibinfo {pages} {206--212} (\bibinfo {year} {2016})}\BibitemShut
  {NoStop}%
\bibitem [{\citenamefont {Stamps}(2014)}]{stamps2014artificial}%
  \BibitemOpen
  \bibfield  {author} {\bibinfo {author} {\bibfnamefont {R.~L.}\ \bibnamefont
  {Stamps}},\ }\href@noop {} {\bibfield  {journal} {\bibinfo  {journal} {Nature
  Phys.}\ }\textbf {\bibinfo {volume} {10}},\ \bibinfo {pages} {623} (\bibinfo
  {year} {2014})}\BibitemShut {NoStop}%
\bibitem [{\citenamefont {Gilbert}\ \emph {et~al.}(2016)\citenamefont
  {Gilbert}, \citenamefont {Lao}, \citenamefont {Carrasquillo}, \citenamefont
  {O’Brien}, \citenamefont {Watts}, \citenamefont {Manno}, \citenamefont
  {Leighton}, \citenamefont {Scholl}, \citenamefont {Nisoli},\ and\
  \citenamefont {Schiffer}}]{gilbert2016emergent}%
  \BibitemOpen
  \bibfield  {author} {\bibinfo {author} {\bibfnamefont {I.}~\bibnamefont
  {Gilbert}}, \bibinfo {author} {\bibfnamefont {Y.}~\bibnamefont {Lao}},
  \bibinfo {author} {\bibfnamefont {I.}~\bibnamefont {Carrasquillo}}, \bibinfo
  {author} {\bibfnamefont {L.}~\bibnamefont {O’Brien}}, \bibinfo {author}
  {\bibfnamefont {J.~D.}\ \bibnamefont {Watts}}, \bibinfo {author}
  {\bibfnamefont {M.}~\bibnamefont {Manno}}, \bibinfo {author} {\bibfnamefont
  {C.}~\bibnamefont {Leighton}}, \bibinfo {author} {\bibfnamefont
  {A.}~\bibnamefont {Scholl}}, \bibinfo {author} {\bibfnamefont
  {C.}~\bibnamefont {Nisoli}}, \ and\ \bibinfo {author} {\bibfnamefont
  {P.}~\bibnamefont {Schiffer}},\ }\href@noop {} {\bibfield  {journal}
  {\bibinfo  {journal} {Nature Phys.}\ }\textbf {\bibinfo {volume} {12}},\
  \bibinfo {pages} {162} (\bibinfo {year} {2016})}\BibitemShut {NoStop}%
\bibitem [{\citenamefont {Arava}\ \emph {et~al.}(2019)\citenamefont {Arava},
  \citenamefont {Leo}, \citenamefont {Schildknecht}, \citenamefont {Cui},
  \citenamefont {Vijayakumar}, \citenamefont {Derlet}, \citenamefont
  {Kleibert},\ and\ \citenamefont {Heyderman}}]{arava2019engineering}%
  \BibitemOpen
  \bibfield  {author} {\bibinfo {author} {\bibfnamefont {H.}~\bibnamefont
  {Arava}}, \bibinfo {author} {\bibfnamefont {N.}~\bibnamefont {Leo}}, \bibinfo
  {author} {\bibfnamefont {D.}~\bibnamefont {Schildknecht}}, \bibinfo {author}
  {\bibfnamefont {J.}~\bibnamefont {Cui}}, \bibinfo {author} {\bibfnamefont
  {J.}~\bibnamefont {Vijayakumar}}, \bibinfo {author} {\bibfnamefont {P.~M.}\
  \bibnamefont {Derlet}}, \bibinfo {author} {\bibfnamefont {A.}~\bibnamefont
  {Kleibert}}, \ and\ \bibinfo {author} {\bibfnamefont {L.~J.}\ \bibnamefont
  {Heyderman}},\ }\href@noop {} {\bibfield  {journal} {\bibinfo  {journal}
  {Phys. Rev. Appl.}\ }\textbf {\bibinfo {volume} {11}},\ \bibinfo {pages}
  {054086} (\bibinfo {year} {2019})}\BibitemShut {NoStop}%
\bibitem [{\citenamefont {Keswani}\ and\ \citenamefont
  {Das}(2019)}]{keswani2019micromagnetic}%
  \BibitemOpen
  \bibfield  {author} {\bibinfo {author} {\bibfnamefont {N.}~\bibnamefont
  {Keswani}}\ and\ \bibinfo {author} {\bibfnamefont {P.}~\bibnamefont {Das}},\
  }\href@noop {} {\bibfield  {journal} {\bibinfo  {journal} {J. Appl. Phys.}\
  }\textbf {\bibinfo {volume} {126}},\ \bibinfo {pages} {214304} (\bibinfo
  {year} {2019})}\BibitemShut {NoStop}%
\bibitem [{\citenamefont {Luo}\ \emph {et~al.}(1988)\citenamefont {Luo},
  \citenamefont {Ohno}, \citenamefont {Matsuzaki},\ and\ \citenamefont
  {Hasegawa}}]{luo1988low}%
  \BibitemOpen
  \bibfield  {author} {\bibinfo {author} {\bibfnamefont {J.-K.}\ \bibnamefont
  {Luo}}, \bibinfo {author} {\bibfnamefont {H.}~\bibnamefont {Ohno}}, \bibinfo
  {author} {\bibfnamefont {K.}~\bibnamefont {Matsuzaki}}, \ and\ \bibinfo
  {author} {\bibfnamefont {H.}~\bibnamefont {Hasegawa}},\ }\href@noop {}
  {\bibfield  {journal} {\bibinfo  {journal} {Jpn. J. Appl. Phys.}\ }\textbf
  {\bibinfo {volume} {27}},\ \bibinfo {pages} {1831} (\bibinfo {year}
  {1988})}\BibitemShut {NoStop}%
\bibitem [{\citenamefont {G{\"o}kta{\c{s}}}\ \emph {et~al.}(2008)\citenamefont
  {G{\"o}kta{\c{s}}}, \citenamefont {Weber}, \citenamefont {Weis},\ and\
  \citenamefont {von Klitzing}}]{goktacs2008alloyed}%
  \BibitemOpen
  \bibfield  {author} {\bibinfo {author} {\bibfnamefont {O.}~\bibnamefont
  {G{\"o}kta{\c{s}}}}, \bibinfo {author} {\bibfnamefont {J.}~\bibnamefont
  {Weber}}, \bibinfo {author} {\bibfnamefont {J.}~\bibnamefont {Weis}}, \ and\
  \bibinfo {author} {\bibfnamefont {K.}~\bibnamefont {von Klitzing}},\
  }\href@noop {} {\bibfield  {journal} {\bibinfo  {journal} {Physica E}\
  }\textbf {\bibinfo {volume} {40}},\ \bibinfo {pages} {1579--1581} (\bibinfo
  {year} {2008})}\BibitemShut {NoStop}%
\bibitem [{\citenamefont {Keswani}\ \emph {et~al.}(2018)\citenamefont
  {Keswani}, \citenamefont {Nakajima}, \citenamefont {Chauhan}, \citenamefont
  {Kumar}, \citenamefont {Ohno},\ and\ \citenamefont
  {Das}}]{keswani2018fabrication}%
  \BibitemOpen
  \bibfield  {author} {\bibinfo {author} {\bibfnamefont {N.}~\bibnamefont
  {Keswani}}, \bibinfo {author} {\bibfnamefont {Y.}~\bibnamefont {Nakajima}},
  \bibinfo {author} {\bibfnamefont {N.}~\bibnamefont {Chauhan}}, \bibinfo
  {author} {\bibfnamefont {S.}~\bibnamefont {Kumar}}, \bibinfo {author}
  {\bibfnamefont {H.}~\bibnamefont {Ohno}}, \ and\ \bibinfo {author}
  {\bibfnamefont {P.}~\bibnamefont {Das}},\ }\href@noop {} {\bibfield
  {journal} {\bibinfo  {journal} {AIP Conf. Proc.}\ }\textbf {\bibinfo {volume}
  {1953}},\ \bibinfo {pages} {050071} (\bibinfo {year} {2018})}\BibitemShut
  {NoStop}%
\bibitem [{\citenamefont {Keswani}\ and\ \citenamefont
  {Das}(2018)}]{keswani2018magnetization}%
  \BibitemOpen
  \bibfield  {author} {\bibinfo {author} {\bibfnamefont {N.}~\bibnamefont
  {Keswani}}\ and\ \bibinfo {author} {\bibfnamefont {P.}~\bibnamefont {Das}},\
  }\href@noop {} {\bibfield  {journal} {\bibinfo  {journal} {AIP Adv.}\
  }\textbf {\bibinfo {volume} {8}},\ \bibinfo {pages} {101501} (\bibinfo {year}
  {2018})}\BibitemShut {NoStop}%
\bibitem [{\citenamefont {Novoselov}\ \emph {et~al.}(2003)\citenamefont
  {Novoselov}, \citenamefont {Geim}, \citenamefont {Dubonos}, \citenamefont
  {Hill},\ and\ \citenamefont {Grigorieva}}]{novoselov2003subatomic}%
  \BibitemOpen
  \bibfield  {author} {\bibinfo {author} {\bibfnamefont {K.}~\bibnamefont
  {Novoselov}}, \bibinfo {author} {\bibfnamefont {A.}~\bibnamefont {Geim}},
  \bibinfo {author} {\bibfnamefont {S.}~\bibnamefont {Dubonos}}, \bibinfo
  {author} {\bibfnamefont {E.}~\bibnamefont {Hill}}, \ and\ \bibinfo {author}
  {\bibfnamefont {I.}~\bibnamefont {Grigorieva}},\ }\href@noop {} {\bibfield
  {journal} {\bibinfo  {journal} {Nature}\ }\textbf {\bibinfo {volume} {426}},\
  \bibinfo {pages} {812} (\bibinfo {year} {2003})}\BibitemShut {NoStop}%
\bibitem [{\citenamefont {Lipert}\ \emph {et~al.}(2010)\citenamefont {Lipert},
  \citenamefont {Bahr}, \citenamefont {Wolny}, \citenamefont {Atkinson},
  \citenamefont {Wei{\ss}ker}, \citenamefont {M{\"u}hl}, \citenamefont
  {Schmidt}, \citenamefont {B{\"u}chner},\ and\ \citenamefont
  {Klingeler}}]{lipert2010individual}%
  \BibitemOpen
  \bibfield  {author} {\bibinfo {author} {\bibfnamefont {K.}~\bibnamefont
  {Lipert}}, \bibinfo {author} {\bibfnamefont {S.}~\bibnamefont {Bahr}},
  \bibinfo {author} {\bibfnamefont {F.}~\bibnamefont {Wolny}}, \bibinfo
  {author} {\bibfnamefont {P.}~\bibnamefont {Atkinson}}, \bibinfo {author}
  {\bibfnamefont {U.}~\bibnamefont {Wei{\ss}ker}}, \bibinfo {author}
  {\bibfnamefont {T.}~\bibnamefont {M{\"u}hl}}, \bibinfo {author}
  {\bibfnamefont {O.}~\bibnamefont {Schmidt}}, \bibinfo {author} {\bibfnamefont
  {B.}~\bibnamefont {B{\"u}chner}}, \ and\ \bibinfo {author} {\bibfnamefont
  {R.}~\bibnamefont {Klingeler}},\ }\href@noop {} {\bibfield  {journal}
  {\bibinfo  {journal} {Appl. Phys. Lett.}\ }\textbf {\bibinfo {volume} {97}},\
  \bibinfo {pages} {212503} (\bibinfo {year} {2010})}\BibitemShut {NoStop}%
\bibitem [{\citenamefont {Das}\ \emph {et~al.}(2010)\citenamefont {Das},
  \citenamefont {Porrati}, \citenamefont {Wirth}, \citenamefont {Bajpai},
  \citenamefont {Huth}, \citenamefont {Ohno}, \citenamefont {Ohno},\ and\
  \citenamefont {M{\"u}ller}}]{das2010magnetization}%
  \BibitemOpen
  \bibfield  {author} {\bibinfo {author} {\bibfnamefont {P.}~\bibnamefont
  {Das}}, \bibinfo {author} {\bibfnamefont {F.}~\bibnamefont {Porrati}},
  \bibinfo {author} {\bibfnamefont {S.}~\bibnamefont {Wirth}}, \bibinfo
  {author} {\bibfnamefont {A.}~\bibnamefont {Bajpai}}, \bibinfo {author}
  {\bibfnamefont {M.}~\bibnamefont {Huth}}, \bibinfo {author} {\bibfnamefont
  {Y.}~\bibnamefont {Ohno}}, \bibinfo {author} {\bibfnamefont {H.}~\bibnamefont
  {Ohno}}, \ and\ \bibinfo {author} {\bibfnamefont {J.}~\bibnamefont
  {M{\"u}ller}},\ }\href@noop {} {\bibfield  {journal} {\bibinfo  {journal}
  {Appl. Phys. Lett.}\ }\textbf {\bibinfo {volume} {97}},\ \bibinfo {pages}
  {042507} (\bibinfo {year} {2010})}\BibitemShut {NoStop}%
\bibitem [{\citenamefont {Matsunaga}\ \emph {et~al.}(2013)\citenamefont
  {Matsunaga}, \citenamefont {Furukawa}, \citenamefont {Kanda}, \citenamefont
  {Hara}, \citenamefont {Nomura},\ and\ \citenamefont
  {Kimura}}]{matsunaga2013detection}%
  \BibitemOpen
  \bibfield  {author} {\bibinfo {author} {\bibfnamefont {T.}~\bibnamefont
  {Matsunaga}}, \bibinfo {author} {\bibfnamefont {K.}~\bibnamefont {Furukawa}},
  \bibinfo {author} {\bibfnamefont {Y.}~\bibnamefont {Kanda}}, \bibinfo
  {author} {\bibfnamefont {M.}~\bibnamefont {Hara}}, \bibinfo {author}
  {\bibfnamefont {T.}~\bibnamefont {Nomura}}, \ and\ \bibinfo {author}
  {\bibfnamefont {T.}~\bibnamefont {Kimura}},\ }\href@noop {} {\bibfield
  {journal} {\bibinfo  {journal} {Appl. Phys. Lett.}\ }\textbf {\bibinfo
  {volume} {102}},\ \bibinfo {pages} {252405} (\bibinfo {year}
  {2013})}\BibitemShut {NoStop}%
\bibitem [{\citenamefont {Donahue}\ and\ \citenamefont
  {Porter}(1999)}]{donahue1999national}%
  \BibitemOpen
  \bibfield  {author} {\bibinfo {author} {\bibfnamefont {M.}~\bibnamefont
  {Donahue}}\ and\ \bibinfo {author} {\bibfnamefont {D.}~\bibnamefont
  {Porter}},\ }\bibfield  {title} {\enquote {\bibinfo {title} {National
  institute of standards and technology interagency report no},}\ }\href@noop
  {} {\bibfield  {journal} {\bibinfo  {journal} {NISTIR6376}\ } (\bibinfo
  {year} {1999})}\BibitemShut {NoStop}%
\bibitem [{\citenamefont {Guimar{\~a}es}(2009)}]{principles}%
  \BibitemOpen
  \bibfield  {author} {\bibinfo {author} {\bibfnamefont {A.~P.}\ \bibnamefont
  {Guimar{\~a}es}},\ }\href@noop {} {\emph {\bibinfo {title} {Principles of
  nanomagnetism}}}\ (\bibinfo  {publisher} {Springer- Berlin},\ \bibinfo {year}
  {2009})\BibitemShut {NoStop}%
\end{thebibliography}%

\end{document}